\documentclass[runningheads]{llncs}
\usepackage{graphicx}
\usepackage{color}
\usepackage{xcolor}
\definecolor{qb}{rgb}{1,0.5,0}
\begin{document}
%
\title{U-Net Training with Instance-Layer Normalization}
\author{Xiao-Yun Zhou\inst{1} \and Peichao Li\inst{1} \and Zhao-Yang Wang\inst{1} \and Guang-Zhong Yang\inst{1,2}}
\authorrunning{X.-Y. Zhou et al.}
\institute{The Hamlyn Centre for Robotic Surgery, Imperial College London, UK \and Institute of Medical Robotics, Shanghai Jiao Tong University, China \email{xiaoyun.zhou14@imperial.ac.uk}}
%
%
%
\maketitle              
\begin{abstract}
Normalization layers are essential in a Deep Convolutional Neural Network (DCNN). Various normalization methods have been proposed. The statistics used to normalize the feature maps can be computed at batch, channel, or instance level. However, in most of existing methods, the normalization for each layer is fixed. Batch-Instance Normalization (BIN) is one of the first proposed methods that combines two different normalization methods and achieve diverse normalization for different layers. However, two potential issues exist in BIN: first, the Clip function is not differentiable at input values of 0 and 1; second, the combined feature map is not with a normalized distribution which is harmful for signal propagation in DCNN. In this paper, an Instance-Layer Normalization (ILN) layer is proposed by using the Sigmoid function for the feature map combination, and cascading group normalization. The performance of ILN is validated on image segmentation of the Right Ventricle (RV) and Left Ventricle (LV) using U-Net as the network architecture. The results show that the proposed ILN outperforms previous traditional and popular normalization methods with noticeable accuracy improvements for most validations, supporting the effectiveness of the proposed ILN.
\keywords{Instance-Layer Normalization \and Deep Convolutional Neural Network \and U-Net \and Biomedical Image Segmentation.}
\end{abstract}
\section{Introduction}
Biomedical image segmentation is a fundamental step in medical image analysis, i.e., 3D shape instantiation for organs~\cite{zhou2018real} and prosthesis~\cite{zhou2018towards,zhou2018realstent}. Most current popular methods are based on Deep Convolutional Neural Network (DCNN) which train multiple non-linear modules for feature extraction and pixel classification with both higher automation and performance. One fundamental component in DCNN is the normalization layer. Initially, one of the main motivations for normalization was to alleviate the internal covariate shift where layers' input distribution changes~\cite{ioffe2015batch}. However, recent work considers the use of normalization layer is beneficial, because it increases the robustness of the networks to fluctuation associated with random initialization~\cite{bjorck2018understanding}, or it achieves smoother optimization landscape~\cite{santurkar2018does}. In this paper, we keep this motivation question open and focus on normalization strategies.

For a feature map with dimension of $(\rm N, H, W, C)$, where $\rm N$ is the batch size, $\rm H$ is the feature height, $\rm W$ is the feature width, $\rm C$ is the feature channel, Batch Normalization (BN)~\cite{ioffe2015batch}\cite{ioffe2017batch} was the first proposed normalization method which calculated the mean and variance of a feature map along the $(\rm N, H, W)$ dimension, then re-scaled and re-translated the normalized feature map with additional trainable parameters to preserve the DCNN representation ability. Instance Normalization (IN)~\cite{ulyanov2016instance} which calculated the mean and variance along the $(\rm H, W)$ dimension was proposed for fast stylization. Layer Normalization (LN)~\cite{ba2016layer} which calculated the mean and variance along the $(\rm H, W, C)$ dimension was proposed for recurrent networks. Group Normalization (GN)~\cite{wu2018group} calculated the mean and variance along the $(\rm H, W)$ and multiple-channels dimension and was validated on image classification and instance segmentation. A review of these four normalization methods for training U-Net for medical image segmentation could be found in~\cite{zhou2019normalization}. Weight normalization~\cite{salimans2016weight}\cite{xu2018understanding} based on re-parameterization on weights was used in recurrent models and reinforcement learning. Batch Kalman normalization estimated the mean and variance considering all preceding layers~\cite{wang2018batch}. 

Recently, Nam \textit{et al.} proposed Batch-Instance Normalization~\cite{nam2018batch} (BIN), which combined BN and IN with a trainable parameter. However, two risks potentially exist: 1) the trainable parameter was restricted in the range of [0, 1] with Clip function which is not differentiable at input values of 0 and 1; 2) the combined feature map was no longer with a normalized distribution, which is harmful for signal propagation in DCNN. In this paper, Instance-Layer Normalization (ILN) is proposed to combine IN and LN: 1) Sigmoid is used to solve the non-differentiable characteristic of Clip function at input values of 0 and 1; 2) an additional GN16 - GN with a group number of 16 is added after the combined feature map to ensure a normalized distribution of the combined feature map. A widely-applied and popular network architecture - U-Net ~\cite{ronneberger2015u} is used as the network to validate the proposed ILN on the Right Ventricle (RV) and Left Ventricle (LV) image segmentation. The proposed ILN outperforms existing normalization methods with noticeable accuracy improvements in most validations in terms of the Dice Similarity Coefficient (DSC).
\section{Methodology}
\subsection{Instance-Layer Normalization}
\subsubsection{Instance Normalization}
\begin{figure}
\centering
\includegraphics[width=0.5\textwidth]{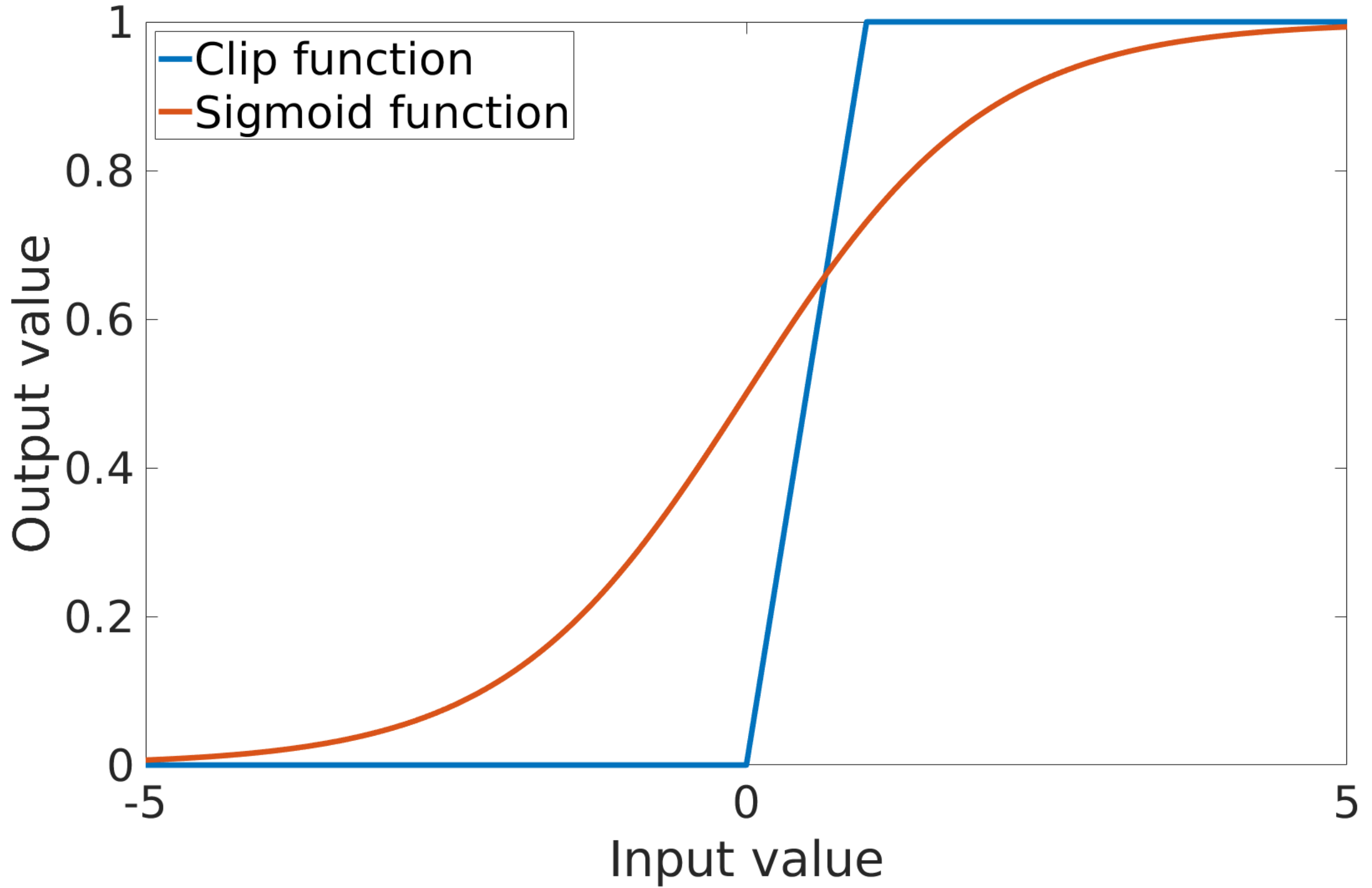}
\caption{The curves of Clip and Sigmoid function.}
\label{fig:Clip_sigmoid}
\end{figure}

With a feature map $\textbf{F}$ of dimension $\rm (N, H, W, C)$, IN calculates the mean and variance of $\textbf{F}$ as:
\begin{equation}
\mu_{n,c} = \frac{1}{\rm H \times \rm W} \sum\limits_{h=1}^{\rm H} \sum\limits_{w=1}^{\rm W} f_{n,h,w,c};\quad
\delta_{n,c}^2 = \frac{1}{\rm H \times \rm W} \sum\limits_{h=1}^{\rm H} \sum\limits_{w=1}^{\rm W} (f_{n,h,w,c}-\mu_{n,c})^2
\end{equation}
Then, the feature map is normalized as $\hat{\textbf{F}}^{\rm I}$:
\begin{equation}
\label{equ:norm}
\hat{f}_{n,h,w,c}^{\rm I} = \frac{f_{n,h,w,c}-\mu_{n,c}}{\sqrt{\delta_{n,c}^2+\epsilon}}
\end{equation}
where $\epsilon$ is a small value added for division stability. For the same feature map $\textbf{F}$, LN calculates the mean and variance as:
\begin{equation}
\mu_n = \frac{1}{\rm H \times \rm W \times \rm C } \sum\limits_{h=1}^{\rm H} \sum\limits_{w=1}^{\rm W} \sum\limits_{c=1}^{\rm C} f_{n,h,w,c};
\delta_n^2 = \frac{1}{\rm H \times \rm W \times \rm C} \sum\limits_{h=1}^{\rm H} \sum\limits_{w=1}^{\rm W} \sum\limits_{c=1}^{\rm C} (f_{n,h,w,c}-\mu_n)^2
\end{equation}
where $\textbf{F}$ is normalized in a similar way of Equ. (\ref{equ:norm}) to $\hat{\textbf{F}}^{\rm L}$. A trainable parameter $\rho$ is added to combine $\hat{\textbf{F}}^I$ and $\hat{\textbf{F}}^{\rm L}$. In the original BIN~\cite{nam2018batch}, $\rho$ was clipped to be in the range of $[0, 1]$ with a Clip function, as shown in Figure~\ref{fig:Clip_sigmoid}.

\noindent
However, Clip function is not differentiable at input values of $0$ and $1$. In this paper, Sigmoid function $Sigmoid(x)=1/(e^{-x}+1)$ which is differentiable everywhere is applied to solve this potential issue:
\begin{equation}
\hat{\textbf{F}}^{\rm IL} = Sigmoid( \rho ) \cdot \hat{\textbf{F}}^{\rm I} + (1-Sigmoid( \rho )) \cdot \hat{\textbf{F}}^{\rm L}
\end{equation}
An additional potential issue in the original BIN is that the combined $\hat{\textbf{F}}^{\rm IL}$ is no longer with a mean of $0$ and a variance of $1$, this non-normalized distribution may be harmful for signal propagation in DCNN. In this paper, we solve this issue with applying an additional GN16 on the combined $\hat{\textbf{F}}^{\rm IL}$:
\begin{equation}
\mu_{n,g} = \frac{1}{\rm H \times \rm W \times \rm M} \sum\limits_{h=1}^{\rm H} \sum\limits_{w=1}^{\rm W} \sum\limits_{m=(g-1)\cdot M+1}^{\rm g\cdot M} \hat{f}_{n,h,w,m}^{\rm IL}, \rm M=C//16
\end{equation}
\begin{equation}
\delta_{n,g}^2 = \frac{1}{\rm H \times \rm W \times \rm M} \sum\limits_{h=1}^{\rm H} \sum\limits_{w=1}^{\rm W} \sum\limits_{m=(g-1)\cdot M+1}^{\rm g\cdot M} (\hat{f}_{n,h,w,m}^{\rm IL}-\mu_{n,g})^2, \rm M=C//16
\end{equation}
where $\rm M$ is the channel number in each feature group, // is exact division, $g\in[1,16]$. The feature map is normalized in a similar way of Equ. (\ref{equ:norm}) as $\hat{\textbf{F}}^{ILN}$. Following BN~\cite{ioffe2015batch}, additional parameters $\gamma$ and $\beta$ are added to preserve the DCNN representation ability $f'^{ILN}_{n,h,w,c} = \gamma_c \hat{f}_{n,h,w,c}^{ILN}+\beta_c$.
\subsection{Experimental Setup}
\paragraph{Network Architecture}
A widely adopted network architecture in medical image segmentation, called U-Net~\cite{ronneberger2015u}, was used as the fundamental network framework with four max-pooling layers. The start feature channel number is 16. The normalization layer was added between the convolutional and Relu layer. Cross-entropy was used as the loss function. Momentum Stochastic Gradient Descent (SGD) was used as the optimizer with the momentum set as 0.9. Weights were initialized with a truncated normal distribution with the \textit{stddev} as $2/(3^2 \times \rm C)$, where $\rm C$ is the channel number. Biases were initialized as 0.1. $\rho$ was initialized as 0.5.
\paragraph{Data Collections}
6082 RV images~\cite{zhou2018real}, scanned with a 1.5T Magnetic Resonance Imaging (MRI) machine (Sonata, Siemens, Erlangen, Germany), with slice gap of 10mm, pixel spacing of 1.5$\sim$2mm, image size of $256\times256$, from 37 subjects mixed with Hypertrophic Cardiomyopathy (HCM) patients and asymptomatic subjects, from the atrioventricular ring to the apex were used for the validation. The ground truth was labeled by one expert with Analyze (AnalyzeDirect, Inc, Overland Park, KS, USA). $[-30^\circ:10^\circ:30^\circ]$ rotations were applied to augment the training images. 12, 12, 13 subjects for each group were split randomly for three-fold cross validation. 805 LV images~\cite{radau2009evaluation}, from SunnyBrook MRI dataset, with subject number of 45, image size of $256\times256$, were used for the validation as well. $[-60^\circ:2^\circ:60^\circ]$ rotations were applied to augment the training images. 15 subjects for each group were split randomly for three-fold cross validation.
\paragraph{Implementation}
As the proposed ILN needs to manipulate intermediate feature maps, the U-Net framework was implemented with low-level Tensorflow functions - tf.nn. In this paper, to ensure a fair comparison, all normalization methods were re-implemented into the same framework as the ILN implementation instead of using the available high-level Tensorflow Application Programming Interface (API) exists for some normalization methods in Tensorflow library, such as those used in~\cite{zhou2019normalization}.
\paragraph{Experiments}
Following~\cite{zhou2019normalization} and~\cite{zhou2019atrous}, two epochs were trained for each experiment with dividing the learning rate by 5 at the second epoch. Five initial learning rates $(1.5, 1.0, 0.5, 0.1, 0.05)$ were tested for each experiment and the best value was selected to be shown. DSC was used as the evaluation metric.
\section{Result}
To prove the advantage of using the Sigmoid function over the Clip function (in original BIN \cite{nam2018batch}), three comparison experiments were set up: 1) using Clip function with one trainable parameter $Clip(\rho)_0^1$ for IN feature map while the parameter for LN feature map is $1-Clip(\rho)_0^1$; 2) using Sigmoid function with one trainable parameter $Sigmoid(\rho)$ for IN feature map while the parameter for LN feature map is $1-Sigmoid(\rho)$; 3) using Softmax function with two trainable parameters $Softmax(\rho_1, \rho_2)$ for IN and LN feature map respectively. Comparison results are shown in Section \ref{result:Sigmoid}.

To prove the advantage of adding GN16 after the combined feature map, two comparison experiments with or without GN16 are conducted. Results are shown in Section \ref{result:GN16}. Eight randomly-selected segmentation examples are shown in Section \ref{result:example} for intuitive illustrations. As GN16 performed similarly to IN~\cite{zhou2019normalization}, no normalization, IN, LN, GN4 are chosen as the baseline to validate the performance of the proposed ILN, as presented in details in Section \ref{result:AN}. The training curves of $\rho$ at eight randomly-selected layers are shown in Section \ref{result:rho}. In this paper, RV-1 refers to the cross validation that uses the first group of RV data as testing while uses the second and third group of RV data as training. Similar fashions were applied as the notations of the experiments on the RV-2, RV-3, LV-1, LV-2, and LV-3.
\subsection{Sigmoid vs. Clip vs. Softmax Function}
\label{result:Sigmoid}
The mean$\pm$std segmentation DSCs of using Clip, Sigmoid and Softmax function to combine the IN and LN feature map are shown in Table \ref{tab:clip}. We can see that Sigmoid function achieves the highest DSC for most cross validations, except RV-1 experiment, which proves the effectiveness of the proposed method in this paper - replacing the Clip function in original BIN \cite{nam2018batch} with Sigmoid function.
\begin{table}
\caption{Mean$\pm$std segmentation DSCs of using Clip, Sigmoid and Softmax function to combine the feature map of IN and LN, highest DSCs are in blue colour.}
\label{tab:clip}
\begin{tabular}{|l|l|l|l|l|l|l|}
\hline
Method & RV-1 & RV-2 & RV-3 & LV-1 & LV-2 & LV-3 \\
\hline
Clip & \textcolor{blue}{0.702$\pm$0.295} & 0.707$\pm$0.299 & 0.666$\pm$0.319 & 0.900$\pm$0.099 & 0.864$\pm$0.184 & 0.804$\pm$0.246 \\
Sigmoid & 0.692$\pm$0.304 & \textcolor{blue}{0.724$\pm$0.284} & \textcolor{blue}{0.675$\pm$0.301} & \textcolor{blue}{0.903$\pm$0.118} & \textcolor{blue}{0.888$\pm$0.135} & \textcolor{blue}{0.828$\pm$0.189} \\
Softmax & 0.688$\pm$0.290 & 0.720$\pm$0.279 & 0.664$\pm$0.323 & 0.895$\pm$0.151 & 0.866$\pm$0.153 & 0.827$\pm$0.228 \\
\hline
\end{tabular}
\end{table}
\subsection{With or Without GN16}
\label{result:GN16}
The mean$\pm$std segmentation DSCs of adding or not adding GN16 after the combined feature map of IN and LN are shown in Table \ref{tab:GN}. We can see that, the method with adding GN16 achieves the highest DSC for most cross validations, except LV-3 experiment. This result proves the effectiveness of adding GN16 after the combined feature map and also proves the importance of maintaining the normalized distribution of feature maps.
\begin{table}
\caption{Mean$\pm$std segmentation DSCs of adding or not adding GN16 after the combined feature map of IN and LN, highest DSCs are in blue colour.}
\label{tab:GN}
\begin{tabular}{|l|l|l|l|l|l|l|}
\hline
Method & RV-1 & RV-2 & RV-3 & LV-1 & LV-2 & LV-3 \\
\hline
No & 0.692$\pm$0.304 & 0.724$\pm$0.284 & 0.675$\pm$0.301 & 0.903$\pm$0.118 & 0.888$\pm$0.135 & \textcolor{blue}{0.828$\pm$0.189} \\
Yes & \textcolor{blue}{0.714$\pm$0.290} & \textcolor{blue}{0.737$\pm$0.267} & \textcolor{blue}{0.680$\pm$0.305} & \textcolor{blue}{0.919$\pm$0.098} & \textcolor{blue}{0.893$\pm$0.127} & 0.827$\pm$0.211 \\
\hline
\end{tabular}
\end{table}
\subsection{Segmentation Examples}
\label{result:example}
\begin{figure}
\centering
\includegraphics[width=1.0\textwidth]{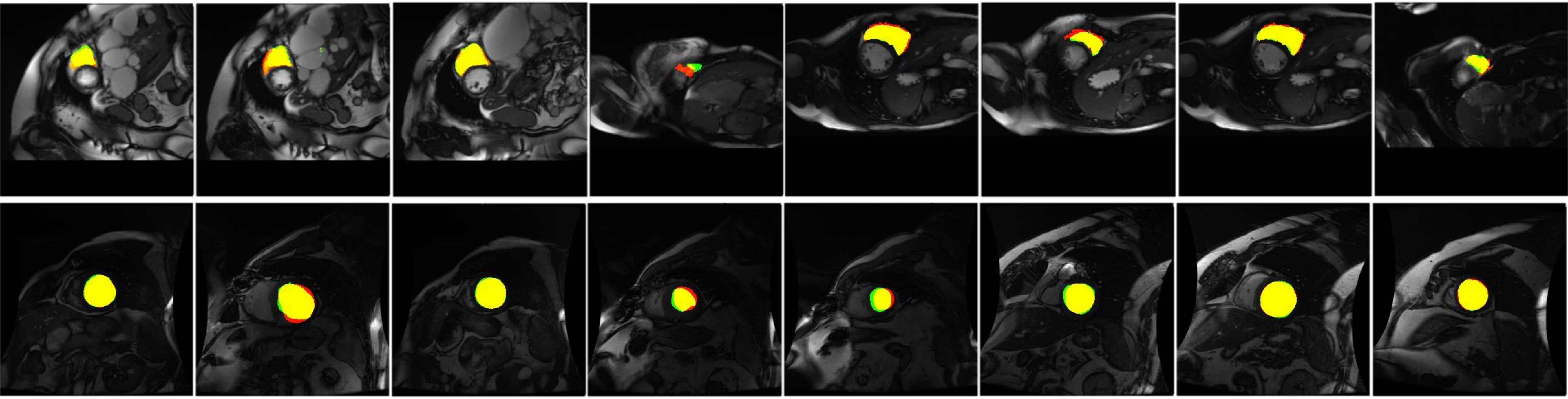}
\caption{Eight examples were selected randomly from the RV and LV segmentation results, where red indicates the ground truth, green indicates the segmentation result, and yellow indicates the true positives of the prediction.}
\label{fig:example}
\end{figure}

Eight segmentation examples were selected randomly from the RV and LV data to show the segmentation details in Figure \ref{fig:example}. For most cases, both the RV and LV are segmented properly. However, for cases near the RV apex, i.e., the forth figure in the first row, the segmentation quality is worse. This might be due to the tissue adhesion and the small size of RV.
\subsection{Comparison to Other Methods}
\label{result:AN}
The mean$\pm$std segmentation DSCs of using no normalization, IN, LN, GN4, and the proposed ILN with the U-Net framework are shown in Table \ref{tab:comparison}. We can see that, except the LV-3 experiment, the proposed ILN outperforms all other traditional methods with considerable accuracy improvements. This result proves the effectiveness of the proposed ILN in medical image segmentation.
\begin{table}
\caption{Mean$\pm$std segmentation DSCs of using no normalization, IN, LN, GN4, and the proposed ILN with the U-Net framework, highest DSCs are in blue colour.}
\label{tab:comparison}
\begin{tabular}{|l|l|l|l|l|l|l|}
\hline
Method & RV-1 & RV-2 & RV-3 & LV-1 & LV-2 & LV-3 \\
\hline
None & 0.688$\pm$0.296 & 0.678$\pm$0.318 & 0.661$\pm$0.323 & 0.899$\pm$0.134 & 0.872$\pm$0.167 & 0.784$\pm$0.280 \\
IN & 0.709$\pm$0.266 & 0.715$\pm$0.278 & 0.655$\pm$0.327 & 0.905$\pm$0.114 & 0.876$\pm$0.131 & \textcolor{blue}{0.836$\pm$0.207} \\
LN & 0.702$\pm$0.287 & 0.718$\pm$0.270 & 0.662$\pm$0.309 & 0.898$\pm$0.120 & 0.858$\pm$0.187 & 0.793$\pm$0.262 \\
GN4 & 0.679$\pm$0.303 & 0.701$\pm$0.291 & 0.671$\pm$0.309 & 0.908$\pm$0.113 & 0.841$\pm$0.196 & 0.800$\pm$0.255 \\
ILN & \textcolor{blue}{0.714$\pm$0.290} & \textcolor{blue}{0.737$\pm$0.267} & \textcolor{blue}{0.680$\pm$0.305} & \textcolor{blue}{0.919$\pm$0.098} & \textcolor{blue}{0.893$\pm$0.127} & 0.827$\pm$0.211 \\
\hline
\end{tabular}
\end{table}
\subsection{Training Curves of $\rho$}
\label{result:rho}
\begin{figure}
\centering
\includegraphics[width=1.0\textwidth]{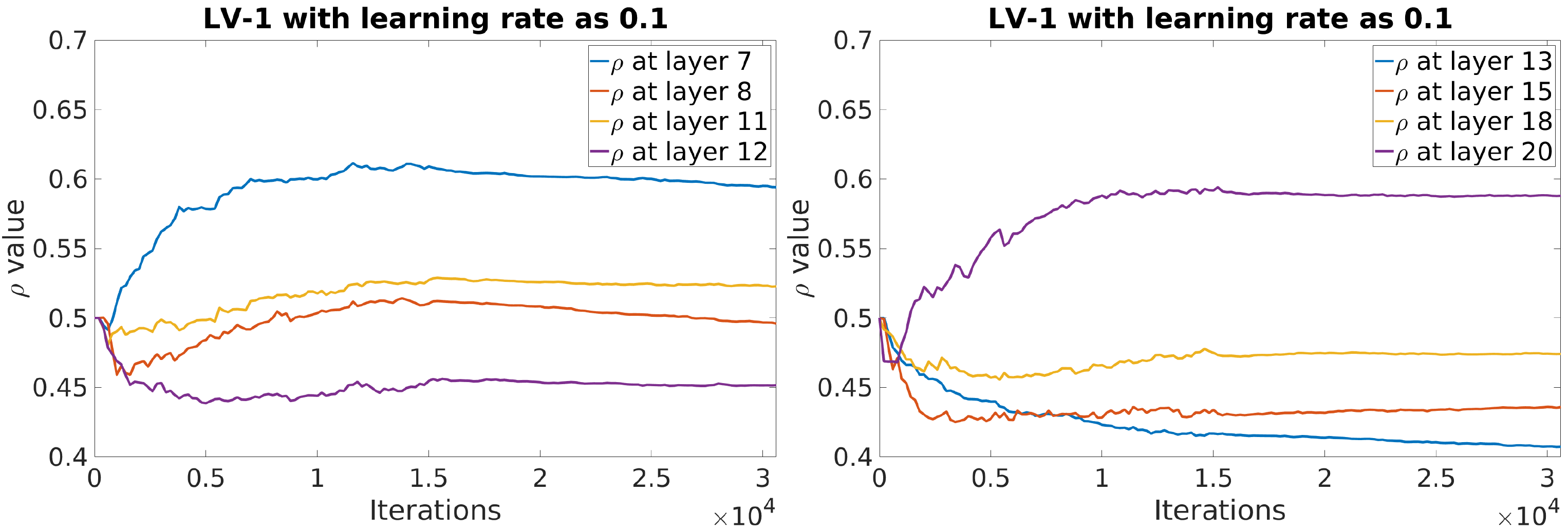}
\caption{The training curves of eight $\rho$ selected randomly from the 22 layers in U-Net.}
\label{fig:patient_errir}
\end{figure}

The $\rho$ training curves of eight layers were selected randomly from LV-1 experiment to be shown in Figure \ref{fig:patient_errir}. We can see that $\rho$ was trained to be different values and the proposed ILN achieved diverse normalization at different layers. As the ground truth of $\rho$ is not known and it is impossible to judge the curve correctness, a comparison regarding the $\rho$ training curves of ILN and BIN is not illustrated.

The CPU used is Intel Xeon(R) E5-1650 v4@3.60GHz$\times$12. The GPU used is Nvidia Titan XP. Comparing ILN to IN, the parameter number increases 22, as one parameter is added to each layer. The training time for 200 iterations increases from 34.8s to 36.5s due to the additional GN16 calculation.
\section{Discussion}
The proposed ILN strategy is generic and flexible. The three components, IN, LN and GN16 could be replaced with other normalization methods. The proposed ILN framework is validated on medical image segmentation with a U-Net framework. We believe that it could also be useful for other tasks, which needs further validation and exploration. The proposed ILN failed to achieve the highest DSC for the LV-3 experiment. It may due to that the combination of IN, LN and GN16 is not suitable for this experiment. In the future, the proposed ILN framework would be extended to combining more normalization methods.
\section{Conclusion}
To improve the accuracy of biomedical image segmentation based on U-net, the ILN was proposed to combine the feature map of IN and LN with an additional trainable parameter and Sigmoid function, then add GN16 after the combined feature map. Although, various normalization methods have been proposed, the noticeable accuracy improvements of the proposed ILN - almost $2\%$ DSC proves the importance of carefully tuning the normalization strategy when training DCNNs.
\bibliographystyle{splncs04}
\bibliography{Adaptivenorm}

\begin{thebibliography}{10}
\providecommand{\url}[1]{\texttt{#1}}
\providecommand{\urlprefix}{URL }
\providecommand{\doi}[1]{https://doi.org/#1}

\bibitem{ba2016layer}
Ba, J.L., Kiros, J.R., Hinton, G.E.: Layer normalization. Stat  \textbf{1050},
  ~21 (2016)

\bibitem{bjorck2018understanding}
Bjorck, N., Gomes, C.P., Selman, B., Weinberger, K.Q.: Understanding batch
  normalization. In: NeurIPS. pp. 7705--7716 (2018)

\bibitem{ioffe2017batch}
Ioffe, S.: Batch renormalization: Towards reducing minibatch dependence in
  batch-normalized models. In: NeurIPS. pp. 1945--1953 (2017)

\bibitem{ioffe2015batch}
Ioffe, S., Szegedy, C.: Batch normalization: Accelerating deep network training
  by reducing internal covariate shift. In: ICML. pp. 448--456 (2015)

\bibitem{nam2018batch}
Nam, H., Kim, H.E.: Batch-instance normalization for adaptively style-invariant
  neural networks. In: NeurIPS. pp. 2563--2572 (2018)

\bibitem{radau2009evaluation}
Radau, P., Lu, Y., Connelly, K., Paul, G., Dick, A., Wright, G.: Evaluation
  framework for algorithms segmenting short axis cardiac {MRI}. The MIDAS
  Journal-Cardiac MR Left Ventricle Segmentation Challenge  \textbf{49} (2009)

\bibitem{ronneberger2015u}
Ronneberger, O., Fischer, P., Brox, T.: U-net: Convolutional networks for
  biomedical image segmentation. In: MICCAI. pp. 234--241 (2015)

\bibitem{salimans2016weight}
Salimans, T., Kingma, D.P.: Weight normalization: A simple reparameterization
  to accelerate training of deep neural networks. In: NeurIPS. pp. 901--909
  (2016)

\bibitem{santurkar2018does}
Santurkar, S., Tsipras, D., Ilyas, A., Madry, A.: How does batch normalization
  help optimization? In: NeurIPS. pp. 2488--2498 (2018)

\bibitem{ulyanov2016instance}
Ulyanov, D., Vedaldi, A., Lempitsky, V.: Instance normalization: The missing
  ingredient for fast stylization. arXiv preprint arXiv:1607.08022  (2016)

\bibitem{wang2018batch}
Wang, G., Peng, J., Luo, P., Wang, X., Lin, L.: Batch kalman normalization:
  Towards training deep neural networks with micro-batches. arXiv preprint
  arXiv:1802.03133  (2018)

\bibitem{wu2018group}
Wu, Y., He, K.: Group normalization. In: ECCV. pp. 3--19 (2018)

\bibitem{xu2018understanding}
Xu, Y., Wang, X.: Understanding weight normalized deep neural networks with
  rectified linear units. In: NeurIPS. pp. 130--139 (2018)

\bibitem{zhou2018realstent}
Zhou, X.Y., Lin, J., Riga, C., Yang, G.Z., Lee, S.L.: Real-time {3D} shape
  instantiation from single fluoroscopy projection for fenestrated stent graft
  deployment. IEEE RAL  \textbf{3}(2),  1314--1321 (2018)

\bibitem{zhou2018towards}
Zhou, X.Y., Riga, C., Lee, S.L., Yang, G.Z.: Towards automatic {3D} shape
  instantiation for deployed stent grafts: {2D} multiple-class and
  class-imbalance marker segmentation with equally-weighted focal {U-Net}. In:
  2018 IEEE/RSJ IROS. pp. 1261--1267 (2018)

\bibitem{zhou2019normalization}
Zhou, X.Y., Yang, G.Z.: Normalization in training {U-Net} for {2D} biomedical
  semantic segmentation. IEEE RAL  (2019)

\bibitem{zhou2018real}
Zhou, X.Y., Yang, G.Z., Lee, S.L.: A real-time and registration-free framework
  for dynamic shape instantiation. MedIA  \textbf{44},  86--97 (2018)

\bibitem{zhou2019atrous}
Zhou, X.Y., Zheng, J.Q., Yang, G.Z.: Atrous convolutional neural network
  ({ACNN}) for biomedical semantic segmentation with dimensionally lossless
  feature maps. arXiv preprint arXiv:1901.09203  (2019)

\end{thebibliography}
\end{document}